\documentclass[twocolumn,showpacs,preprintnumbers,amsmath,amssymb]{revtex4}
\usepackage{graphicx}
\usepackage{setspace}
\newcommand{\ket}[1]{|#1\rangle}

\begin{document}
\title{Electronic structure and magnetic properties of $\mathrm{CaCr}\mathrm{O}_3$: The interplay between spin- and orbital-orderings}
\author{Wei Wu}\email{wei.wu@ucl.ac.uk}
\affiliation{Department of Electronic and Electrical Engineering and London Centre for Nanotechnology, University College London, Gower Street, London, WC1E 6BT, United Kingdom}


\begin{abstract}
The electronic structure and magnetic properties of CaCr$\mathrm{O}_3$ have been calculated by two methods, including hybrid-exchange density-function theory and density-functional theory + $U$. The computed densities of states from both of these methods are in a qualitative agreement with the previous x-ray spectroscopy. On the other hand, the opening of the band gap separates them apart. hybrid-exchange density-functional theory always gives a finite band gap, down to $\sim 1.2$ eV from HSE06 functional, whereas by tuning the Hubbard-$U$ parameter down to $0.5$ eV, a conducting state with AFM-C (defined in the text) spin configuration can be achieved. From hybrid density-functional theory, the computed nearest-neighbouring exchange interaction along the $c$-axis and in the $ab$-plane are $\sim 4$ meV and $\sim 6$ meV (anti-ferromagnetic), respectively, which are qualitatively in agreement with the previous magnetic measurements. The density-functional theory + $U$ has predicted a similar series of exchange interactions. These anti-ferromagnetic exchange interaction, together with the in-plane anti-ferro-orbital ordering will induce a spin-orbital frustration, which could play a role for the abnormal electronic properties in CaCrO$_3$. In hybrid-exchange density-functional theory, an abrupt reduction ($\sim 0.2$ eV) of the majority-spin band gap of the ferromagnetic state between 60 K and 100 K has been found as lowering temperature, which shows a strong link to the previous optical conductivity measurements in [A. C. Komarek, et. al., Phys. Rev. B \textbf{84}, 125114 (2011)]. In sharp contrast, the density-functional theory + $U$ methods predicted AFM-C state as the lowest AFM state for the crystal structure measured below 90 K, above which AFM-A is however the lowest. The closely related concepts including electron-hole liquid and surface-plasmon-mediating spin-spin interactions have been discussed as well.
\end{abstract}


\maketitle

\section{Introduction}




Oxides, widely known as ceramics in the modern world, can be even found in the ancient pottery. Among these, transition-metals oxides (TMOs) have copious interesting physical phenomena, ranging from high-transition-temperature superconductors \cite{ybco}, colossal magnetoresistance \cite{ramirez1997}, and multiferroicity \cite{cheong2007} to plasmonics \cite{naik2012}. For example, chromium dioxide (Cr$\mathrm{O}_2$), which is a half metal (conducting in one spin channel but insulating in the other),  has a great potential for spintronics \cite{cro2}, thus exploring electron spins to store and transport information. These fascinating physical phenomena, as well as their important applications in the development of new technologies, hugely depend on the understanding of the interaction between (i) the localised $d$-electron of transition metals and $p$-electron of oxygen and (ii) between the spin and orbital degrees of freedom. Moreover, covalent, metallic, and ionic bonds can all be found in TMOs, hence leading to possible boundary crossing between different electronic and magnetic states. The underlying physics behind these interactions are also manifested in the phenomena of fundamental interest, such as spin- and orbital-orderings. 

\begin{figure}[htbp]
\centering
\includegraphics[scale=0.3,clip=true]{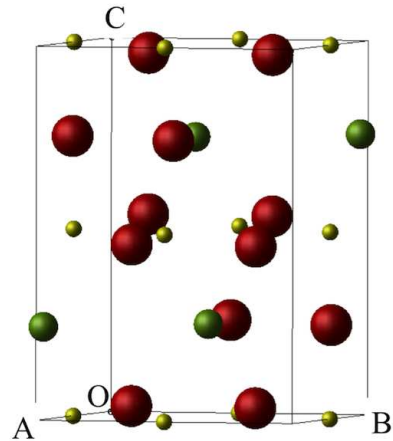}
\caption{ (Colour on line.) The conventional cell of perovskite CaCr$\mathrm{O}_3$ is shown. Ca is depicted as green ball, O as red, and Cr as yellow.}\label{fig:cacro3structure}
\end{figure}

$\mathrm{Cr}^{4+}$-based perovskites, such as CaCr$\mathrm{O}_3$, SrCr$\mathrm{O}_3$, and PbCr$\mathrm{O}_3$, have attracted considerable attentions recently \cite{zhou2006, komarek2008, streltsov2008, bhobe2011, liu2011} owing to their peculiar electronic properties. Especially in CaCr$\mathrm{O}_3$ (whose crystal structure is shown in Fig.\ref{fig:cacro3structure}), the complicated interplay between magnetism, conductivities, and lattice vibrations has been the major obstacle for a clear observation of its electronic ground state \cite{zhou2006, komarek2008}. The question, whether CaCr$\mathrm{O}_3$ is metallic or insulating, is still under active debate. A comprehensive experimental study has suggested that the anomalous properties of CaCr$\mathrm{O}_3$ are related to the Cr-O bond instabilities owing to the spin and orbital orderings, which could result in a crossover from localised to itinerant behaviour \cite{zhou2006}. However, a more resent study, combining experiments and first-principles calculations, has shown that CaCr$\mathrm{O}_3$ is a rare case of anti-ferromagnetic metal owing to the co-existence of anti-ferromagnetism and metallic-like conductivity \cite{komarek2008, bhobe2011}. The theoretical work therein replying on local spin density approximation (LSDA) + $U$ supported that the $c$-type anti-ferromagnetic (AFM) state, in which the spins are in ferromagnetic (FM) ordering along the $c$-axis but AFM in the $ab$-plane, is the ground state \cite{streltsov2008}. the electronic state was predicted to be metallic within LSDA, but insulating once an appropriate value of $U$ was included. Another set of first-principles calculations \cite{liu2011} based on generalised gradient approximation (GGA) + $U$ have suggested that the $c$-type AFM is a metallic state with an intermediate $U$. To the author's knowledge, most of the previous theoretical work has been done either within pure density-functional theory (DFT) or DFT + $U$ \cite{streltsov2008, liu2011} for only one crystal structure. Against this background, Komarek, et. al., have measured the crystal structures in a wide range of temperatures from $T = 3.5 K$ to $300 K$. This work has provided a good opportunity to study the behavior of the electronic structures as a function of temperature, which is yet to be studied thoroughly.  

In this paper, the electronic structure of CaCr$\mathrm{O}_3$ was computed within two theoretical frameworks: (i) hybrid-exchange density-functional theory (HDFT), PBE0 \cite{adamo1998} and HSE06 \cite{krukau2006}, both of which are hybrid-exchange density functional free of any adjustable parameter, and (ii) DFT + $U$ methods where the Hubbard-$U$ can be adjusted, according to different crystal structure reported in \cite{cm2008, komarek2011}. The total energies of ferromagnetic (FM), A-type AFM (AFM-A, AFM along the $c$-axis but FM in the $ab$ plane), C-type AFM (AFM-C), and G-type AFM (N\'eel state, AFM-G) states were computed carefully. The nearest-neighbouring (NN) exchange interactions along $c$-axis and in the $ab$-plane were then extracted by comparing the total energies of different magnetic states. We have found that AFM-G spin configuration is the ground state, which is different from that observed and predicted in Ref.\cite{komarek2008}. However, the total energy difference between the AFM-G and AFM-C states is $< 30$ meV within the reach of room-temperature. These calculations suggest that spin- and orbital- orderings co-exist in CaCrO$_3$. The rest of the discussion is organised as the following: in \S\ref{sec:computationaldetails} the computational details are given, in \S\ref{sec:resultsanddiscussions} the calculation results are presented and discussed, and in \S\ref{sec:conclusion} some more general conclusions are drawn.  

\section{Computational details}\label{sec:computationaldetails}
\subsection{Hybrid-exchange density-functional theory}
Calculations for the electronic structures and magnetic properties of Cr$\mathrm{O}_2$ and CaCr$\mathrm{O}_3$ were carried out by using DFT and a variety of approximate exchange-correlation functionals, including PBE0 \cite{adamo1998}, HSE06 \cite{krukau2006}, and HISS \cite{henderson2007}, implemented in the CRYSTAL14 code \cite{crystal14}.  The experimentally determined lattice parameters and atomic coordinates are adopted here \cite{cm2008, komarek2011}. The basis sets of Ca \cite{valenzano2006}, Cr \cite{catti1996}, and O \cite{towler1994} that are developed for solid-state calculations were used. The Monkhorst-Pack samplings \cite{packmonkhorst} of reciprocal space are carried out choosing a grid of shrinking factor to be $7\times 7 \times 5$ to be consistent with the ratios among reciprocal lattice parameters. The truncation of the Coulomb and exchange series in direct space is controlled by setting the Gaussian overlap tolerance criteria to $10^{-6}, 10^{-6}, 10^{-6}, 10^{-6}$, and $10^{-12}$ \cite{crystal14}. The self-consistent field (SCF) procedure is converged to a tolerance of $10^{-6}$ a.u. per unit cell (p.u.c). To accelerate convergence of the SCF process, all calculations have been performed adopting a linear mixing of Fock matrices by $30$\%.

Electronic exchange and correlation have been described using the PBE0 hybrid-exchange, short- (HSE06), and middle (HISS)-range-corrected functionals. Apart from a partial elimination of the self-interaction error, these functionals balance the tendencies to delocalize and localize wave-functions by mixing a quarter of Fock exchange with that from a generalized gradient approximation (GGA) exchange functional \cite{adamo1998}. The performance of the hybrid functional, e.g., B3LYP or PBE0 has previously been shown to provide an accurate description of the electronic structure and magnetic properties for both inorganic and organic compounds \cite{illas2000, muscat2001}. 

\subsection{Density-functional theory + $U$}
The DFT + $U$ method developed in the Quantum Espresso code \cite{qe} has also been used for a benchmark and further comparison with HDFT method. The PBE (Perdew-Burke-Ernzerhof) exchange-correlation functional \cite{pbe}, along with a range of values of Hubbard-U for Cr, has been chosen. In this method, the same set of lattice parameters and atomic coordinates as in HDFT calculations are used for consistency. The Vanderbilt ultrasoft pseudoptentials developed for the PBE exchange-correlation functional have been used for all the elements throughout DFT + $U$ calculations. The same Monkhorst-Pack sampling has been used for DFT + $U$ as the HDFT calculations. The plane0-wave cutoff energies of $\sim 640$ eV and the threshold of SCF energy convergence of $10^{-6}$ eV were employed. A linear mix of 30 \% of the Fock matrix has been used to accelerate the SCF convergence. The electron correlation associated with $3d$ states of Cr are described by DFT + $U$ methods. A range of values of $U = 0.5 , 1.0, 2.0$ and $4.0$ have been chosen to show the transition between metallic and insulating states.

\subsection{spin Hamiltonian}
The Heisenberg model \cite{heisenberg} for the spin-spin interactions in CaCr$\mathrm{O}_3$ is defined here as,
\begin{equation}\label{eq:spinh}
\hat{H}=J_{c}\sum_{ij\in c}{\hat{\vec{S}}_i\cdot\hat{\vec{S}}_j}+J_{ab}\sum_{ij\in ab}{\hat{\vec{S}}_{i}\cdot\hat{\vec{S}}_j},
\end{equation}
where $J_c$ is the exchange interaction between the NN spins along the $c$-axis and $J_{ab}$ between the NN spins in the $ab$ plane respectively. The $i$ and $j$ label the Cr sites. The exchange interactions are determined by
\begin{eqnarray}\label{eq:DeltaE}
J_c&=&(E_{\mathrm{FM}}-E_{a})/8S^2,
\\J_{ab}&=&(E_{\mathrm{FM}}-E_c)/8S^2,
\end{eqnarray}
where $E_{\mathrm{FM}}$, $E_a$, $E_c$, and $E_g$ are the total energies for a conventional cell in the FM, AFM-A, AFM-C, and AFM-G states, respectively. $S=1$ is adopted for a $\mathrm{Cr}^{4+}$ ion. 

\section{Results and discussions}\label{sec:resultsanddiscussions}

\subsection{HDFT calculations}



\subsubsection{Electronic structure of CaCr$\mathrm{O}_3$}

\begin{figure*}[htbp]
\begin{tabular}{cccc}
\centering
&&\\
 \Large{AFM-A}&\Large{AFM-C}&\Large{AFM-G} \\
 &&\\
\includegraphics[scale=0.25]{hse_afma_3d.eps}&\includegraphics[scale=0.25]{hse_afmc_3d.eps}&\includegraphics[scale=0.25]{hse_afmg_3d.eps}\\
 (a)&(b)&(c)\\
&&\\
\includegraphics[scale=0.25]{hse_afma_2p.eps}&\includegraphics[scale=0.25]{hse_afmc_2p.eps}&\includegraphics[scale=0.25]{hse_afmg_2p.eps}\\
(d)&(e)&(f)\\
&&\\
\includegraphics[scale=0.41]{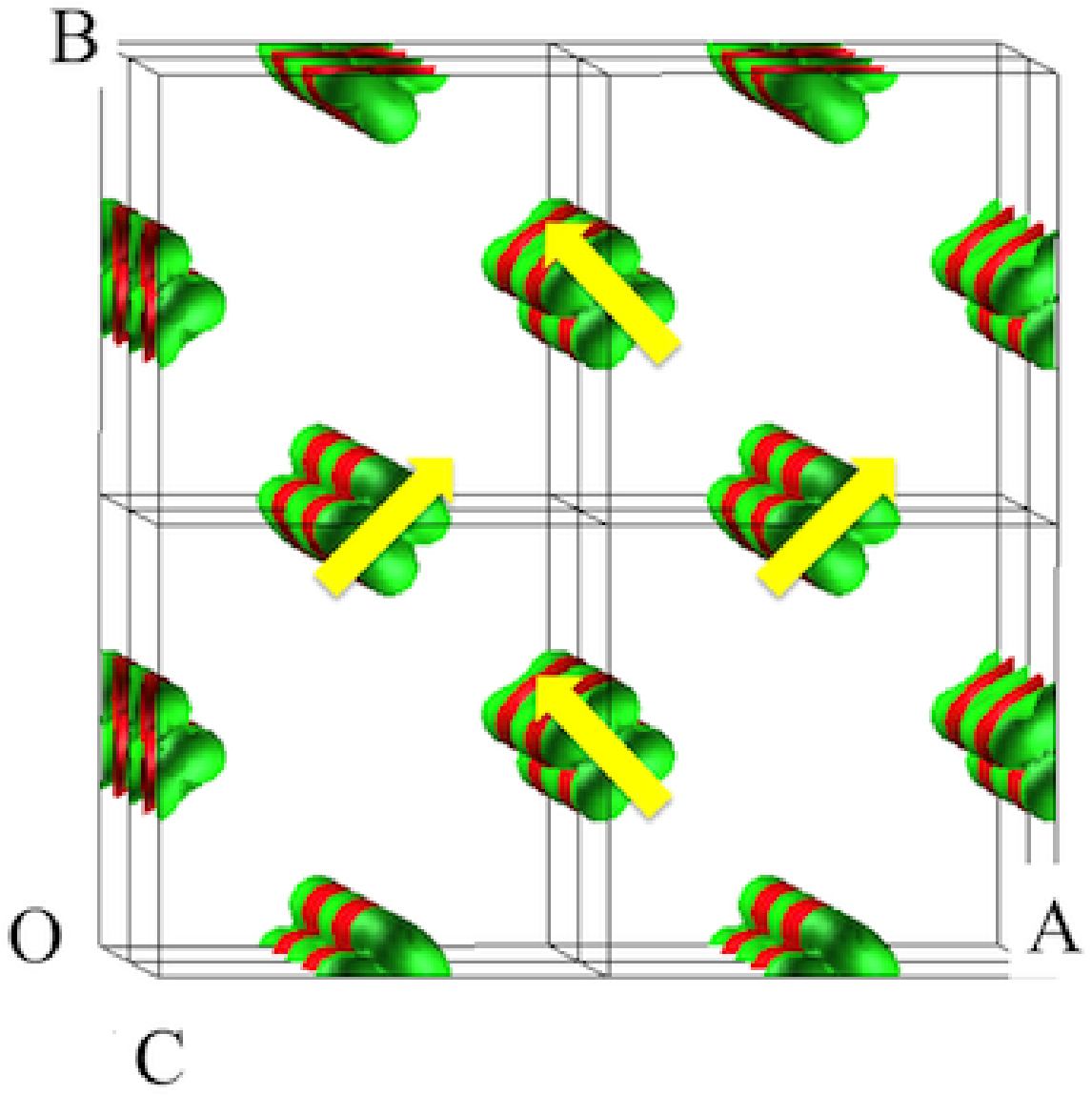}&\includegraphics[scale=0.45]{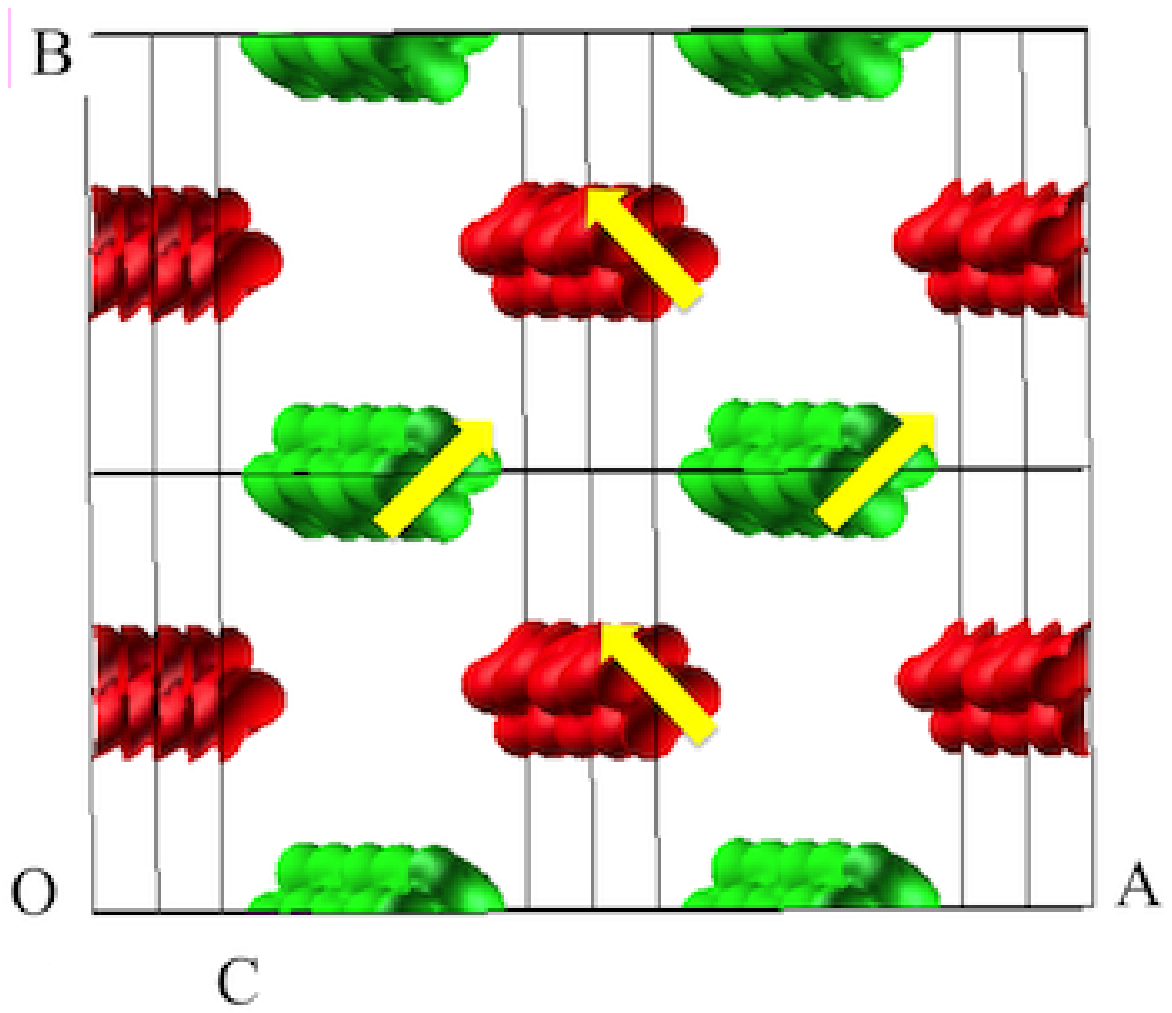}&\includegraphics[scale=0.45]{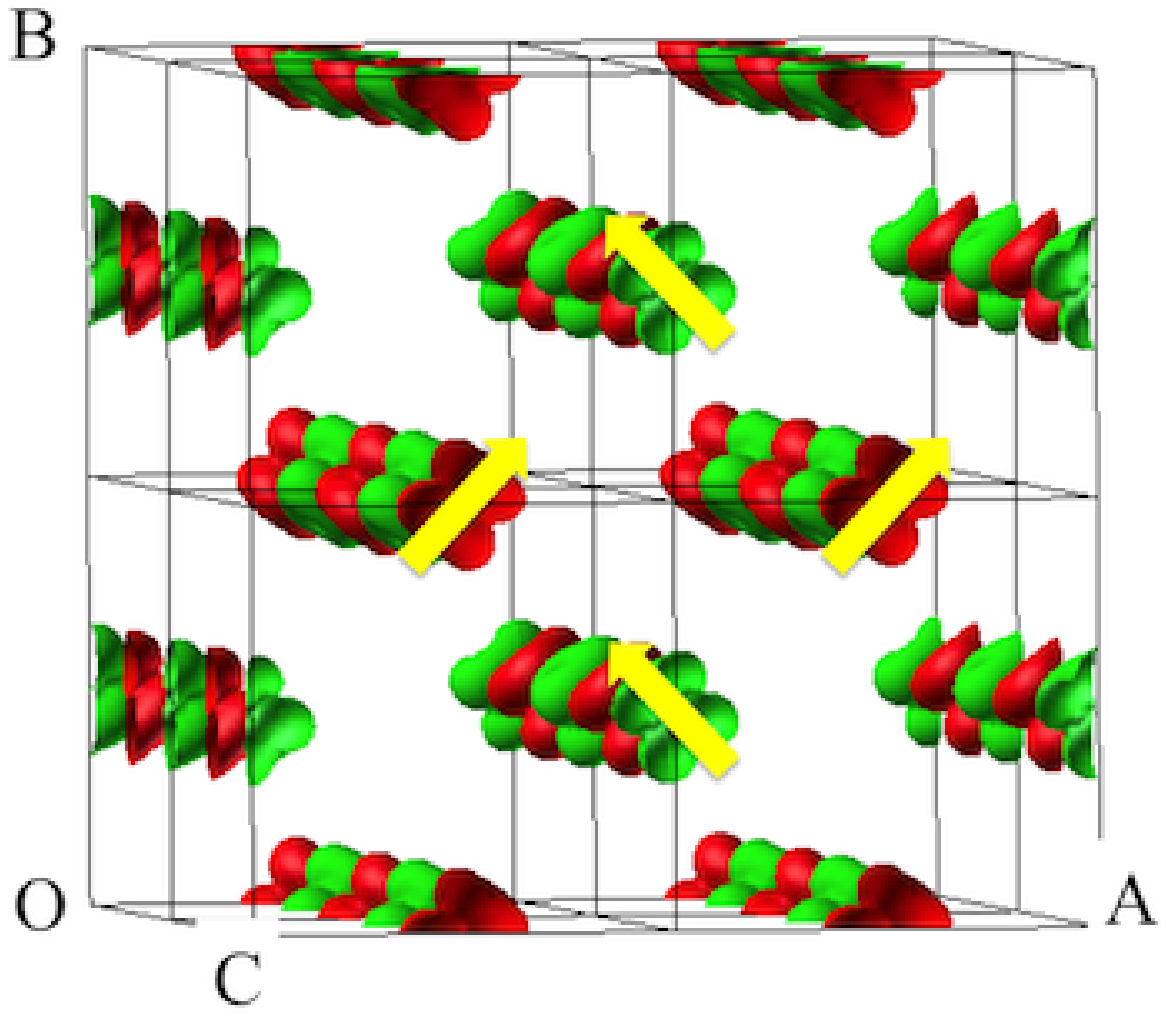}\\
(g)&(h)&(i)\\
\end{tabular}
\caption{ (Colour on line.) The PDOS of Cr $d$-orbitals, O $2p$ orbitals, and spin densities on Cr, computed by using PBE0 for the crystal structure reported in Ref.\cite{cm2008}, are shown for A-type, C-type, and G-type AFM spin configurations for CaCr$\mathrm{O}_3$ in columns. The PDOS for $d_{z^2}$ is depicted in red, $d_{xz}$ in green, $d_{yz}$ in blue, $d_{x^2-y^2}$ in cyan, and $d_{xy}$ in black. The PDOS for $p_x$ is depicted in red, $p_y$ in blue, and $p_z$ in black. Zero-energy is chosen to be at the valence band maximum. In spin densities, spin-up is in red and spin-down in green. Notice that the orbitals are anti-ferro-ordered in the $ab$-plane (the orientation of orbital is illustrated by yellow arrows). }\label{fig:cacro3}
\end{figure*}

The projected density of states (PDOS) onto the $3d$-orbitals of Cr (the first row), $2p$-orbitals of O (the second row), and spin densities (the third row) of the AFM-A, AFM-C, and AFM-G states, calculated by using PBE0 for the crystal structure reported in Ref.\cite{cm2008}, are shown in Fig.\ref{fig:cacro3}. The electronic states of FM, AFM-A, AFM-C, and AFM-G  were predicted to be insulating with a band gap $\sim 2$ eV. The band gaps in different AFM spin configurations, computed by using three functionals, are listed in Table~\ref{tab:dorbitals}. The lowest band gap is given by HSE06 in the AFM-C state, which is $\sim 1.44$ eV for the crystal structure reported in Ref.\cite{cm2008}, while for the crystal structure in Ref.\cite{komarek2011}, the smallest band gap is $\sim 1.2$ eV. Such narrow semiconducting band gap could result in high conductivities and a facile metal-insulator transition. However, PBE0 and HISS have shown slightly larger band gaps, up to $\sim 3$ eV. The AFM-G state is predicted to be the lowest state; this is different from the previous theoretical and experimental work \cite{komarek2008}. This is probably owing to the larger Hubbard-$U$ embedded in the hybrid-exchange-functional methods, as compared to the value of $U$ adopted in the previous local-density approximation + $U$ method \cite{liu2011}. However, the difference between these three AFM states is up to $\sim 38$ meV, which is close to the energy scale of room temperature. Especially the smallest computed energy difference between the AFM-A and AFM-C states is $\sim 2$ meV. Therefore, the stability of these magnetic states is fragile respected to thermal fluctuation and lattice vibration. Thermal and/or lattice fluctuations will further induce a magnetic disorder, which could prevent a proper observation of the electronic state in CaCr$\mathrm{O}_3$. The $3d$-orbital PDOS near VBM is dominated by that from $d_{x^2-y^2}$, $d_{xz}$, and $d_{yz}$. The $2p$-orbital PDOS of O in the AFM spin configurations share the similar feature; one significant peak is near the VBM with the other close to $-3.7$ eV. The comparison between the PDOS of $3d$-orbital and $2p$-orbital suggests there could exist a strong hybridisation between them, which is especially strong near the VBM. This is in agreement with the claim made in the Ref.\cite{bhobe2011} that the electronic properties of CaCrO$_3$ is driven by the strong $2p$-$3d$ hybridisation. This strong bonding between Cr $3d$-orbitals and O $2p$-orbitals might be able to shed some light on the high conductivities of CaCr$\mathrm{O}_3$. Moreover, the O $2p$ PDOS is more dominant than the Cr $3d$ PDOS, which is qualitatively different from CrO$_2$. This could suggest that the conductivity in CaCrO$_3$ originates from predominantly $2p$-orbitals rather than $3d$.

As shown in the spin density of the AFM-A state (Fig.\ref{fig:cacro3}g), the spins are anti-aligned along the $c$-axis, but aligned in the $ab$-plane. In AFM-C (Fig.\ref{fig:cacro3}h), the spins are aligned along the $c$-axis, but anti-aligned in the $ab$-plane. And in AFM-G, a ne\'el state is present, i.e., anti-aligned both along the $c$-axis and in the $ab$-plane. Although there is a clear difference between these magnetic states, they share a common feature that there exists an anti-ferro-orbital ordering (illustrated by yellow arrows) in the $ab$-plane, whereas a ferro-orbital ordering is formed along the $c$-axis. This will result in a frustration between spin- and orbital-orderings in the $ab$-plane for all the AFM states. People have shown in in iron-based superconductors that spin-orbital frustrations that are intrinsically due to quantum fluctuations can induce abnormal metal state \cite{kruger2009}.

\begin{table*}
\begin{tabular}{cccccccccc}
\hline\hline
 &&PBE0&   &&HSE06&   &&HISS& \\
  &AFM-A    &AFM-C    &AFM-G      &AFM-A    &AFM-C    &AFM-G     &AFM-A    &AFM-C    &AFM-G    \\\hline
 $E_{gap} (eV)$ &2.65&2.15&2.50&1.96&1.44&1.89&2.96&2.49&2.97\\ \hline
$d_{z^2}$&0.11&0.11&0.11&0.11&0.11&0.11&0.10&0.10&0.10\\
$d_{xz}$&0.52&0.55&0.51&0.51&0.55&0.52&0.50&0.52&0.49\\
$d_{yz}$&0.54&0.53&0.56&0.55&0.53&0.54&0.54&0.52&0.54\\ 
$d_{x^2-y^2}$&0.60&0.69&0.51&0.70&0.69&0.70&0.72&0.71&0.72\\
$d_{xy}$&0.13&0.14&0.14&0.14&0.14&0.14&0.13&0.13&0.13\\ \hline\hline
\end{tabular}
\caption{The projected spin densities onto Cr $d$-orbitals, computed by using PBE0, HSE06, and HISS are shown for the AFM-A, AFM-C, and AFM-G states.}\label{tab:dorbitals}
\end{table*}
We have also analysed the spin densities projected to $d$-orbitals. As shown in Table \ref{tab:dorbitals}, the spin densities are dominated by $d_{xz}$, $d_{yz}$, and $d_{x^2-y^2}$ orbitals ($\sim 1.7 \mu_B$ for these three orbitals as expected for spin-1) from all the functionals used here. The $d_{x^2-y^2}$ is occupied mainly because the tetragonal distortion, which  will lower $d_{x^2-y^2}$ and lift up the $d_{z^2}$ orbitals. The PDOS calculated here (Fig.\ref{fig:cacro3}) is in a good agreement with this physical picture. The nearly even mixture of $d_{xz}$ and $d_{yz}$ is in agreement with the previous theoretical results that claimed a $t_{2g}$-active orbital ordering \cite{streltsov2008}. The two $d$-orbitals, $d_{xz}$ and $d_{yz}$, are responsible for the complex orbital ordering in the $ab$-plane as shown in Fig.\ref{fig:cacro3}g, h, and i. The orbital ordering in the $ab$-plane is formed by the alternating patterns of the $\frac{1}{\sqrt{2}}(\ket{d_{xz}} + \ket{d_{yz}})$ and $\frac{1}{\sqrt{2}}(\ket{d_{xz}} - \ket{d_{yz}})$ orbitals.

\begin{figure}[htbp]
\centering
\includegraphics[scale=0.45,clip=true]{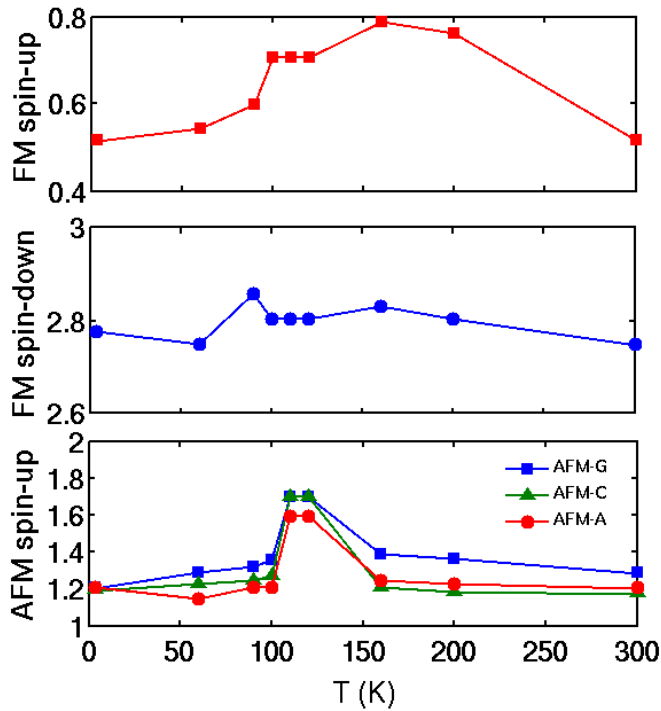}
\caption{ (Colour on line.) DFT band gaps in eV as a function of temperature are shown. The structures are adopted from Ref.\cite{komarek2011}.}\label{fig:fmgap}
\end{figure}

Komarek, et. al. \cite{komarek2011} have observed the evolution of the crystal structure of CaCr$\mathrm{O}_3$ when changing temperatures. The corresponding electronic structures have also been computed by using HSE06 hybrid functional. The DFT band gaps of the majority spins in the FM configuration as a function of temperatures have been plotted in Fig.\ref{fig:fmgap}. The band gap is indirect; the valence band maximum is at the $\Gamma$ point whereas the conduction band minimum (1/2,1/2,0). Notice that there is an abrupt jump of $\sim 0.1$ eV in the FM band gap between 90 K and 100 K, which coincides with the jump in the measurements of resistivity and optical conductivity in Ref. \cite{komarek2011}. In contrast, the minority band gap in FM configuration maintains as $\sim 2.7$ eV. The other band gaps have not been affected by the temperatures as much as FM majority spin.

\subsubsection{Exchange interactions in CaCrO$_3$}
\begin{table}
\begin{tabular}{cccc}
\hline\hline
Exchange (meV) &PBE0&HSE06&HISS \\
$J_c$&4.0&3.3&4.2\\ 
$J_{ab}$&3.1&2.2&4.2\\
\hline\hline
\end{tabular}
\caption{Exchange interactions calculated within different approximations to the exchange-correlation functional for the crystal structure reported in Ref.\cite{cm2008}.}\label{tab:ex1}
\end{table}

The computed exchange interactions for the two sets of crystal structures in Ref.\cite{cm2008} and Ref.\cite{komarek2011} have been tabulated in Table.\ref{tab:ex1} and Table.\ref{tab:ex2}, respectively. Although these exchange interactions are computed within different functionals, they have qualitatively the same structure. The spins are coupled antiferromagnetically along the $c$-axis, but frustrated in the $ab$-plane as the NN exchange interactions is anti-ferromagnetic against the anti-ferro-orbtial ordering. According to Goodenough-Kanamori-Anderson (GKA) rules, this type of orbital ordering should result in a ferromagnetic spin coupling. However, the anti-ferromagnetic NN exchange interaction in the $ab$-plane frustrates this orbital ordering in the $ab$-plane, which might induce a large magnetoelastic coupling \cite{komarek2008}. This could further be linked to the interplay between spin-ordering and lattice vibrations. 

\begin{table}
\begin{tabular}{ccccccccc}
\hline\hline
$T$ (K)&3.5&60&90&100&120&160&200&300\\
$J_c$ (meV)&3.0&2.2&5.8&3.7&9.6&3.4&4.8&2.7\\ 
$J_{ab}$ (meV)&7.2&7.3&5.9&5.6&12.1&6.8&4.9&7.5\\
\hline\hline
\end{tabular}
\caption{Exchange interactions calculated using HSE06 hybrid-exchange functional for the crystal structures reported in Ref.\cite{komarek2011}.}\label{tab:ex2}
\end{table}

\subsection{The DFT + $U$ electronic structure of CaCr$\mathrm{O}_3$}
\begin{figure}[htbp]
\centering
\includegraphics[scale=0.7,clip=true]{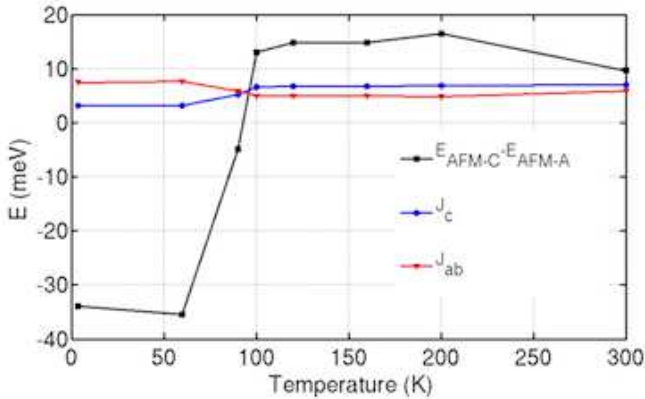}
\caption{ (Colour on line.) The energy differences between AFM-C and AFM-A and exchange interactions computed by using DFT + $U$ method ($U=1$) are shown. The structures are adopted from Ref.\cite{komarek2011}.}\label{fig: dftuenergies}
\end{figure}

We have computed the exchange interactions $J_{ab}$ and $J_c$ and the energy difference between the AFM-A and AFM-C states, according to the DFT + $U$ (with $U=1$) total energies. As shown in Fig.\ref{fig: dftuenergies}, below the N\'eel temperature, the energies of the AFM-A are higher than that of the AFM-C state. The N\'eel temperature is a turning point for the energies difference between the AFM-A and AFM-C states. Correspondingly, this point is important for the magnitude of exchange interactions.

\subsection{Discussions}
\subsubsection{Magnetic states}
The computed PDOS are qualitatively consistent with the x-ray absorption spectroscopy (XAS) as shown in Fig.2a of Ref.\cite{bhobe2011}. The two main peaks in the computed PDOS corroborate the main peak and the shoulder in the O $2p$ intensity of the soft x-ray measurements in Ref.\cite{bhobe2011}. The measured energy gap between O $2p$-orbital and $d$-orbital peaks is $\sim 5$ eV, which is slightly larger than the predicted ($\sim 3.7$ eV). The accurate prediction of the key parameter, Hubbard-$U$, is necessary to understand the nature of the electronic state of CaCrO$_3$, which can be performed by using DFT + $U$ method. To improve further the theoretical description, more advanced computational methods, such as GW \cite{hybertsen1986}, are needed. The inclusion of the spin-orbit coupling (SOC) is another important aspect However, these are beyond the scope of this paper. The N\'eel temperature $T_N$ has been measured from the $1/\chi$ versus $T$ curve, which is $\simeq 90$ K in Ref.\cite{zhou2006, komarek2008}. This is in a good agreement with the computed exchange interactions. In addition, the experimentally observed magnetic moments which were measured to be $3.7 \mu_B$ in Ref.\cite{zhou2006}, which is much larger than the computed ($2.0\  \mu_B$), probably owing to the unsaturated orbital component. However, the observed $1.2 \mu_B$ reported in Ref.\cite{komarek2008} seems an indication of the existence of itinerant magnetism. The inconsistency between the measured magnetic moment could be attributed to the interplay between spin and orbital degrees of freedoms as mentioned previously. 

\subsubsection{Electron-hole liquid}
So far the metallicity in CaCrO$_3$ has predominantly been observed in optical measurements including optical reflectibility and conductivity and photoemissions \cite{komarek2011, bhobe2011}. From the calculations presented here, the FM band gap decreases sharply around $T= 90$ K, meanwhile the indirectness of the band gap has to involve phonon excitations during optical measurements. A closely related concept to the optical conductivity and band gap renormalization is electron-hole liquid (EHL) \cite{beni1978, beni1979}. EHL is a phenomena in which electron-hole density is sufficiently large to effect the electronic structure, i.e., the band gap renormalization. EHL has been observed in the materials with cubic structure and indirect band gaps, and plays a significant role in reducing band gap, leading to a transition from insulator to semimetal \cite{nagai2004}. Recently, it has been shown that the band gap in BaTiO$_3$ nano-particles can be narrowed partially by EHL\cite{ram2014}. Combining with the small gap in the FM configuration as shown in Fig.\ref{fig:fmgap}, at the N\'eel temperature the magnetic fluctuation might play an important role to the formation of EHL in CaCrO$_3$. Based on the calculations presented here and the previous experiments on EHL, we can speculate that the abnormal electronic properties in CaCrO$_3$ could be interpreted as an EHL assisted mainly by magnetic fluctuations, orbital ordering, and small FM band gap. The quasi half-metallicity due to the small FM band gap is crucial for the understanding of the electronic properties of CaCrO$_3$.

\subsubsection{Surface state}
In addition to the usefulness of EHL in plasmon enhancements \cite{ko2013}, the dangling bonds on the surface state in CaCrO$_3$ could be another source of metallicity. On the other hand, presumably the surface state of CaCrO$_3$ is conducting, this would be useful for exciting surface plasmons. In conjugation with the spins carried by $Cr^{4+}$, the CaCrO$_3$ could be a promising candidate for manipulation of spins through surface plasmons \cite{manj2012}.

\section{Conclusions and outlook} \label{sec:conclusion}
In summary, the electronic structure and magnetic properties of $\mathrm{Cr}^{4+}$-based perovskites CaCr$\mathrm{O}_3$ have been calculated within DFT by using hybrid-exchange density functional PBE0, HSE06, and HISS. The computed PDOS is qualitatively in agreement with the previous soft x-ray measurements. The compute exchange interactions are consistent with the previous magnetic measurements. Moreover, the spin densities have shown an anti-ferro-orbital ordering in the $ab$-plane. The analysis of the exchange interactions and orbital ordering shows that there exists a spin-orbital frustration in the $ab$-plane, which could induce abnormal metallic state. The abrupt jump in the band gap of the FM majority spin between 90 K and 100 K corroborates the jump in the optical measurements reported previously. The PDOS have suggested a strong hybridisation between $3d$ and $2p$ orbitals, which agrees with the previous experimental works.  It might be worthwhile performing calculations for the Hubbard-$U$, taking into account SOC and using more advanced computational methods to compute the DOS in the future, which would inspire more discussions about this fascinating material.

\section{Acknowledgement}
I would like to thank Dr Bo Chen and Prof. Nicholas Harrison for stimulating discussions. I am sincerely grateful that Rutherford Appleton Laboratory has kindly provided computational resources. 


\begin{thebibliography}{99}

\bibitem{ybco} M. K. Wu, J. R. Ashburn, C. J. Torng, P. H. Hor, R. L. Meng, L. Gao, Z. J. Huang, Y. Q. Wang, and C. W. Chu,  Phys. Rev. Lett. \textbf{58}, 908 (1987).


\bibitem{ramirez1997} A. P. Ramirez, J. Phys.: Condens. Matter \textbf{9} 8171 (1997).

\bibitem{cheong2007} S. Cheong and M. Mostovoy, Nat. Mat. \textbf{6}, 13 (2007). 

\bibitem{naik2012} G. V. Naik, J. Liu, A. V. Kildishev, V. M. Shalaev, and A. Boltasseva, Proc. Natl. Acad. Sci. \textbf{109}, 8843 (2012).

\bibitem{cro2} J. M. D. Coey and M. Venkatesan, J. App. Phys. \textbf{91}, 8345 (2002).

\bibitem{zhou2006} J.-S. Zhou, C.-Q. Jin, Y.-W. Long, L.-X. Yang, and J. B. Goodenough, Phys. Rev. Lett. \textbf{96}, 046408 (2006).

\bibitem{komarek2008} S. V. Streltsov, M. A. Korotin, V. I. Anisimov, and D. I. Khomskii, Phys. Rev. B \textbf{78}, 054425 (2008).

\bibitem{bhobe2011} P. A. Bhobe, A. Chainani, M. Taguchi, R. Eguchi, M. Matsunami, T. Ohtsuki, K. Ishizaka, M. Okawa, M. Oura, Y. Senba, H. Ohashi, M. Isobe, Y. Ueda, and S. Shin, Phys. Rev. B \textbf{83}, 165132 (2011).

\bibitem{streltsov2008} A. C. Komarek, S.V. Streltsov, M. Isobe, T. M\"oller, M. Hoelzel, A. Senyshyn, D. Trots, M. T. Fern\'andez-Di\'az, T. Hansen, H. Gotou, T. Yagi, Y. Ueda, V. I. Anisimov, M. Gr\"uninger, D. I. Khomskii, and M. Braden, Phys. Rev. Lett. \textbf{101}, 167204 (2008).

\bibitem{liu2011} H. M. Liu, C. Zhu, C. Y. Ma, S. Dong, and J.-M. Liu, J. App. Phys. \textbf{110}, 073701 (2011).

\bibitem{adamo1998} C. Adamo, et. al., J. Chem. Phys., \textbf{110}, 6158 (1998).

\bibitem{krukau2006} A. V. Krukau, O. A. Vydrov, A. F. Izmaylov and G. E. Scuseria, J. Chem. Phys. \textbf{125}, 224106 (2006).

\bibitem{cm2008}E. Castillo-Marti\'nez, A. Dur\'an, and M.\'A . Alario-Franco, J. Sol. Stat. Chem. \textbf{181}, 895 (2008).

\bibitem{komarek2011} A. C. Komarek, T. M\"oller, M. Isobe, Y. Drees, H. Ulbrich, M. Azuma, M. T. Fern\'andez-D\'iaz, A. Senyshyn,
M. Hoelzel, G. Andr\'e, Y. Ueda, M. Gr\'uninger, and M. Braden, Phys. Rev. B \textbf{84}, 125114 (2011).

\bibitem{henderson2007} T. M. Henderson, A. F. Izmaylov, G. E. Scuseria and A. Savin, J. Chem. Phys. \textbf{127}, 221103 (2007).

\bibitem{crystal14} R. Dovesi, V. R. Saunders, C. Roetti, R. Orlando, C. M. Zicovich-Wilson, F. Pascale, B. Civalleri, K. Doll, N. M. Harrison, I. J. Bush, P. D'Arco, M. Llunell, M. Caus\'a and Y. No\"el, CRYSTAL14 User's Manual (University of Torino, Torino, 2014).

\bibitem{valenzano2006} L. Valenzano, F.J. Torres, K. Doll, F. Pascale, C.M. Zicovich-Wilson, R. Dovesi, Z. Phys. Chem. \textbf{220}, 893 (2006).

\bibitem{catti1996} M. Catti, G. Sandrone, G. Valerio and R. Dovesi, J. Phys. Chem. Solids \textbf{57}, 1735 (1996).

\bibitem{towler1994} M. D. Towler, N. L. Allan, N. M. Harrison, V. R. Saunders, W. C. Mackrodt and E. Apra', Phys. Rev. B \textbf{50}, 5041 (1994).

\bibitem{packmonkhorst} H. J. Monkhorst and J. D. Pack, Phys. Rev. B \textbf{13}, 5188 (1976).




\bibitem{heisenberg} W. Heisenberg, Z. Phys. \textbf{49}, 619 (1928).

\bibitem{illas2000} F. Illas, et. al.,Theo. Chem. Acc, \textbf{104}, 265 (2000).

\bibitem{muscat2001} J. Muscata, A. Wanderb, and N.M. Harrison, Chem. Phys. Lett. \textbf{342}, 397 (2001).

\bibitem{qe} P. Giannozzi, S. Baroni, N. Bonini, M. Calandra, R. Car, C. Cavazzoni, D. Ceresoli, G. L. Chiarotti, M. Cococcioni, I. Dabo, A. Dal Corso, S. Fabris, G. Fratesi, S. de Gironcoli, R. Gebauer, U. Gerstmann, C. Gougoussis, A. Kokalj, M. Lazzeri, L. Martin-Samos, N. Marzari, F. Mauri, R. Mazzarello, S. Paolini, A. Pasquarello, L. Paulatto, C. Sbraccia, S. Scandolo, G. Sclauzero, A. P. Seitsonen, A. Smogunov, P. Umari, R. M. Wentzcovitch, J. Phys.: Condens. Matter, \textbf{21}, 395502 (2009).

\bibitem{pbe} J. P. Perdew, K. Burke, and M. Ernzerhof, Phys. Rev. Lett., \textbf{77},  3865 (1996).

\bibitem{kruger2009} F. Kr\"uger, S. Kumar, J. Zaanen, and J. van den Brink, Phys. Rev. B \textbf{79}, 054504 (2009).

\bibitem{gka} J. B. Goodenough, Magnetism and the Chemical Bond (Interscience, New York, 1963).


\bibitem{hybertsen1986} M. S. Hybertsen and S. G. Louie, Phys. Rev. B \textbf{34}, 5390 (1986).

\bibitem{beni1978} G. Beni and T. M. Rice, Phys. Rev. B \textbf{18}, 768 (1978).

\bibitem{beni1979} G. Beni, T. M. Rice, and L. A. Hemstreet, Phys. Rev. B \textbf{19}, 2204 (1979).

\bibitem{nagai2004} T. Nagai, T. J. Inagaki, and Y. Kanemitsu, App. Phys. Lett. \textbf{84}, 1284 (2004).

\bibitem{ram2014} S. Ramakanth and K. C. James Raju, J. App. Phys. \textbf{115}, 173507 (2014).

\bibitem{ko2013} C. Ko, Y. Han, C. Chen, J. Shieh, and M. Chen, J. Phys. Chem. C, \textbf{117}, 26204 (2013).

\bibitem{manj2012} A. Manjavacas, S. Thongrattanasiri, D. E. Chang, and F. J. Garc'a de Abajo, New J. Phys. \textbf{14}, 123020 (2012).

\end{thebibliography}
\end{document}